\documentstyle[epsf]{mn}
\newif\ifAMStwofonts
\ifoldfss
  \ifCUPmtlplainloaded \else
    \NewTextAlphabet{textbfit} {cmbxti10} {}
    \NewTextAlphabet{textbfss} {cmssbx10} {}
    \NewMathAlphabet{mathbfit} {cmbxti10} {} 
    \NewMathAlphabet{mathbfss} {cmssbx10} {} 
  \fi
  \ifAMStwofonts
    \ifCUPmtlplainloaded \else
      \NewSymbolFont{upmath} {eurm10}
      \NewSymbolFont{AMSa} {msam10}
      \NewMathSymbol{\upi}     {0}{upmath}{19}
      \NewMathSymbol{\umu}     {0}{upmath}{16}
      \NewMathSymbol{\upartial}{0}{upmath}{40}
      \NewMathSymbol{\leqslant}{3}{AMSa}{36}
      \NewMathSymbol{\geqslant}{3}{AMSa}{3E}

    \fi
  \fi
\fi 

\ifnfssone
  \newmathalphabet{\mathit}
  \addtoversion{normal}{\mathit}{cmr}{m}{it}
  \addtoversion{bold}{\mathit}{cmr}{bx}{it}
  \newmathalphabet{\mathbfit} 
  \addtoversion{normal}{\mathbfit}{cmr}{bx}{it}
  \addtoversion{bold}{\mathbfit}{cmr}{bx}{it}
  \newmathalphabet{\mathbfss} 
  \addtoversion{normal}{\mathbfss}{cmss}{bx}{n}
  \addtoversion{bold}{\mathbfss}{cmss}{bx}{n}
  \ifAMStwofonts
    \ifCUPmtlplainloaded \else
      \UseAMStwoboldmath
      \makeatletter
      \new@mathgroup\upmath@group
      \define@mathgroup\mv@normal\upmath@group{eur}{m}{n}
      \define@mathgroup\mv@bold\upmath@group{eur}{b}{n}
      \edef\UPM{\hexnumber\upmath@group}
      \new@mathgroup\amsa@group
      \define@mathgroup\mv@normal\amsa@group{msa}{m}{n}
      \define@mathgroup\mv@bold\amsa@group{msa}{m}{n}
      \edef\AMSa{\hexnumber\amsa@group}
      \makeatother
      \mathchardef\upi="0\UPM19
      \mathchardef\umu="0\UPM16
      \mathchardef\upartial="0\UPM40
      \mathchardef\leqslant="3\AMSa36
      \mathchardef\geqslant="3\AMSa3E
    \fi
  \fi
\fi 

\ifnfsstwo
  \DeclareMathAlphabet{\mathbfit}{OT1}{cmr}{bx}{it}
  \SetMathAlphabet\mathbfit{bold}{OT1}{cmr}{bx}{it}
  \DeclareMathAlphabet{\mathbfss}{OT1}{cmss}{bx}{n}
  \SetMathAlphabet\mathbfss{bold}{OT1}{cmss}{bx}{n}
  \ifAMStwofonts
    \ifCUPmtlplainloaded \else
      \DeclareSymbolFont{UPM}{U}{eur}{m}{n}
      \SetSymbolFont{UPM}{bold}{U}{eur}{b}{n}
      \DeclareSymbolFont{AMSa}{U}{msa}{m}{n}
      \DeclareMathSymbol{\upi}{0}{UPM}{"19}
      \DeclareMathSymbol{\umu}{0}{UPM}{"16}
      \DeclareMathSymbol{\upartial}{0}{UPM}{"40}
      \DeclareMathSymbol{\leqslant}{3}{AMSa}{"36}
      \DeclareMathSymbol{\geqslant}{3}{AMSa}{"3E}
    \fi
  \fi
\fi 

\ifCUPmtlplainloaded \else
  \ifAMStwofonts \else 
    \def\upi{\pi}
    \def\umu{\mu}
    \def\upartial{\partial}
  \fi
\fi

\pagerange{\pageref{firstpage}--\pageref{lastpage}}
\pubyear{1999}

\begin{document}
\twocolumn
\title[On the optical pulsations from Geminga]{On the optical pulsations from the Geminga pulsar}
\author[Gil, Khechinashvili \& Melikidze]{Janusz A. Gil$^{1}$, David G. Khechinashvili$^{1,2}$ and
George I. Melikidze$^{1,2}$\\
$^{1}$ J. Kepler Astronomical Center, Pedagogical University, Lubuska 2, 65-265
Zielona G\'ora, Poland\\
$^{2}$ Abastumani Astrophysical Observatory, Al. Kazbegi Avenue 2a, 380060
Tbilisi, Georgia}
\date{Accepted Received in original form}
\maketitle
\begin{abstract}
We present a model for generation mechanisms of the optical pulsations recently detected from the Geminga pulsar. We argue that this is just a synchrotron radiation emitted along open magnetic field lines at altitudes of a few light cylinder radii (which requires that Geminga is an almost aligned rotator), where charged particles acquire non-zero pitch-angles as a result of the cyclotron absorption of radio waves in the magnetized pair plasma (Gil, Khechinashvili \& Melikidze 1998, hereafter Paper I). This explains self-consistently both the lack of apparent radio emission, at least at frequencies higher than about 100~MHz, and the optical pulsations from the Geminga pulsar. From our model it follows that the synchrotron radiation is a maximum in the infrared band, which suggests that Geminga should also be a source of a pulsed infrared emission.
\end{abstract}

\label{firstpage} 
\section{Introduction}
An optical counterpart of the Geminga source was found by Bignami et al. (1987), based upon color considerations. Recent spectral studies (Martin et al. 1998) show a continuous power-law $\left( \nu ^{-0.8\pm 0.5}\right)$ from 3700 to 8000 \AA\ with a broad absorption feature over $6300\div 6500$ \AA~band. Such a spectrum indicates existence of the dominant nonthermal synchrotron component in the optical emission of the Geminga pulsar. 

Recently Shearer et al. (1998), based on their deep integrated images in $B$-band, claimed that the optical counterpart of Geminga pulses with a period $P=237$~ms, reported also at $X$ (Halpern \& Holt 1992) and $\gamma$-ray (Bertch et al. 1992) bands. They derived the magnitude of the pulsed emission  $m_{_{B}}=26.0\pm 0.4$~mag, and found that the signal shows two peaks with a phase separation of $\approx 0.5$. It appeared that there is a phase agreement of the optical data with $\gamma$ and hard $X$-ray curves. Moreover, the form of the optical light curve resembles that of $\gamma$ and hard $X$-rays rather than the soft $X$-rays signature. According to Shearer et al. (1998), this is a further evidence of the predominantly magnetospheric origin of Geminga's pulsed optical emission (Halpern et al. 1996) over a thermal one (Bignami et al. 1996).

We suggest that Geminga's pulsed optical emission is a synchrotron radiation of the plasma particles gyrating about relatively weak magnetic field of the distant magnetosphere of this pulsar. As we demonstrate below, the plasma particles acquire perpendicular momenta, necessary for the synchrotron emission, due to cyclotron absorption of radio waves on these particles. There is an evidence that the Geminga pulsar is either radio quiet or, more likely, visible only at low radio frequencies below about $100$~MHz (e.g. Malofeev \& Malov 1997; Vats et al. 1999;
see also Paper~ I).

\section{Estimation of pitch-angles} \label{Pitch}
Mikhailovskii et al. (1982) were first to suggest that the energy lost by radio waves due to cyclotron damping should lead to 'heating' of the ambient plasma, i.e. increase of its perpendicular temperature. Lyubarskii \& Petrova (1998) treated thoroughly spontaneous re-emission of the absorbed energy, and concluded that in short-period pulsars a
significant fraction of this energy is re-emitted in the far infrared band. They found that an electron (or positron), moving initially with the relativistic velocity strictly along the local magnetic field, obtains a pitch-angle\footnote{The pitch-angle is defined as $\psi \equiv \arctan \left(p_{\perp }/p_{\parallel }\right)$, where $p_{\parallel }$ and $p_{\perp }$ are the components of the particle momentum along and across the local magnetic field, respectively.}
\begin{equation}
\psi \approx \theta \eta ^{1/2} \quad {\rm rad}  \label{psi}
\end{equation}
in the course of cyclotron absorption of radio emission. Here $\theta$ is an angle between the wave vector $\bmath{k}_{\rm{r}}$ and the local magnetic field $\bmath{B}$, and 
\begin{equation}
\eta =\frac{L_{\rm{r}}}{L_{p1}}  \label{eta}
\end{equation}
is the ratio of the radio luminosity to the particle luminosity, that is, the ratio of the power of radio waves (damped due to cyclotron resonance) to the total kinetic energy of the resonant particles produced per second above a fraction of the polar cap with the area $S_{p1}$ (see equation~\ref{S_1}). Note that equation~(\ref{psi}) is valid provided that $\eta\ll 1$. Estimation of $\eta$ is presented below.
\begin{figure}
\epsfxsize=5cm \epsffile[10 26 241 310]{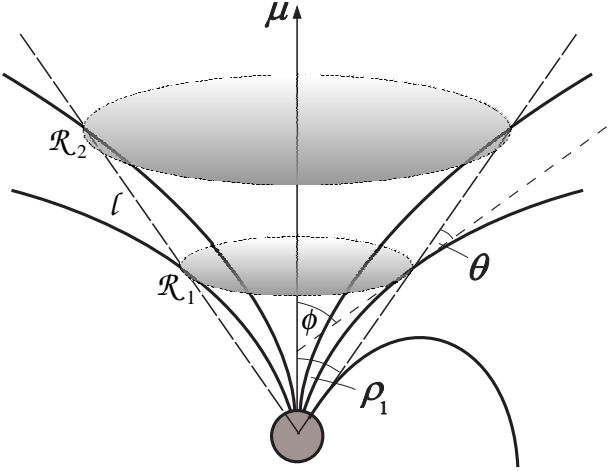}
\caption{Schematic representation of the region where the bulk of Geminga's radio emission is damped. This cut-cone-shaped volume is constrained by the two dashed circles transverse to $\bmu$. The beaming of the optical radiation, re-emitted from this region, is determined by the angle $\phi$. See definitions in the text. \label{Optics region}}
\end{figure}

The radio luminosity of the Geminga pulsar, by what we mean the integrated radio power of Geminga as it would be in the absence of cyclotron damping on the magnetospheric plasma (Paper I), can be estimated as
\begin{equation}
L_{\rm{r}}\approx F\left( 100~\rm{MHz}\right) \,\Delta \nu\,d^{2}\quad {[\rm erg~s^{-1}]},  \label{L_radio}
\end{equation}
where $F\left(100\,{\rm MHz}\right) \approx 1\,{\rm Jy=10^{-23}\,erg~cm^{-2}\,s^{-1} Hz^{-1}}$~(Vats et al. 1999), $\Delta \nu \sim 1$ GHz is a frequency range where the bulk of pulsar radio emission is usually radiated,
and $d\approx 160\,\rm{pc}\approx 5\times 10^{20}$~cm is a distance to the Geminga pulsar. Inserting all these values into equation~(\ref{L_radio}) we find that $L_{\rm r}\approx 3\times 10^{27}~{\rm erg~s^{-1}}$.

Let us now evaluate the denominator in equation~(\ref{eta}). According to Paper I (see also Malov 1998), the Geminga pulsar is an almost aligned rotator, and for the inclination angle $\alpha =5^\circ$ and the impact angle $\beta =20^\circ$, an observer's line of sight in fact grazes the emission cone radiated from the bundle of the last open field lines, with the opening angle at about 100 MHz being $\rho_1\approx 30^\circ$ (Khechinashvili, Melikidze \& Gil 1999). The latter defines a group of the radial lines constituting an angle $\rho_1$ with the magnetic axis $\bmu$ (see Fig.~\ref{Optics region}). The simulation of radio waves damping along these lines (Paper I; Khechinashvili et al. 1999) shows that $\nu \approx 100$~MHz radio waves are damped at the distance ${\mathcal R}_{2}\approx 2.4$, whereas $\nu \approx 1$~GHz waves are damped at ${\mathcal R}_{1}\approx 1.1$. Here ${\mathcal R}\equiv R/R_{_{\rm LC}}$ is a distance from the star's centre measured in the units of the light cylinder radius ($R_{_{\rm LC}}=Pc/2\pi$). One can speculate that the major part of Geminga's radio emission is effectively damped within a cut-cone-shaped volume represented in Fig.~\ref{Optics region}. Then, let us consider the dipolar field line which crosses the radial line, constituting the angle $\rho_1=30^\circ$ with $\bmu$, at the distance ${\mathcal R}_{1}\approx 1.1$. The footprint of this field line on the stellar surface has a polar angle $\vartheta _{p1}\approx \left(R_{0}/R_{1}\right) ^{0.5}\rho_1\approx 0.014$ rad. The footprints of all inner field lines, on which $0.1\div 1$~GHz radio frequencies are damped at high altitudes (see above), mark a circle with the angular radius $\vartheta_{p1}$ within the pulsar polar cap through which the resonant leave the neutron star. Let us notice that $\vartheta_{p1}\approx 0.5\,\vartheta _{p}$, where $\vartheta _{p}\approx (R_0/R_{_{\rm LC}})^{1/2}$ is an angular radius of the pulsar polar cap, and $R_0\approx 10^6$~cm is a stellar radius. Hence, only a fraction of the polar cap provide resonant particles. The area of this region is
\begin{equation}
S_{1}\approx\pi R_{0}^{2}\vartheta _{p1}^{2}\approx 6\times 10^{8}~\rm{cm^{2}}, \label{S_1}
\end{equation}
which yields
\begin{equation}
L_{p1}=\kappa n_{_{\rm{GJ}}}S\gamma_{_p}mc^{3}\approx 8\times 10^{27}\kappa \ 
\left[ \rm{erg~s}^{-1}\right].  \label{L_p}
\end{equation}
Here the surface value of the corotational charge number density (Goldereich \& Julian 1969) $n_{_{\rm{GJ}}}\approx 2.2 \times 10^{18}\left( \dot P/P\right)^{1/2}\approx 5.4\times 10^{12}~{\rm cm^{-3}}$ was substituted, which was evaluated for the dynamical parameters of the Geminga pulsar ($P=0.237$~s and $\dot P=1.1\times 10^{-14}$); $\kappa$ is the Sturrock multiplication factor (Sturrock 1971). Evaluating  equation~(\ref{L_p}) we assumed that the mean Lorentz factor of the ambient plasma particles $\gamma_{_p}\approx 100$. From equation~(\ref{L_p}) it follows that
\begin{equation}
F_{p1}=\frac{L_{p1}}{\gamma_{_p}mc^{2}}\approx 10^{32}\kappa \label{F_p}
\end{equation}
particles are generated per second over the area of $S_{1}$ (equation~\ref{S_1}).

Substituting the derived values $L_{\rm{r}}$ and $L_{p1}$ in equation~(\ref{eta}) we obtain that $\eta \approx 0.4\kappa^{-1}$. Then, taking into account the average value of the angle $\theta \approx 0.6$ rad (Khechinashvili et al. 1999) in equation~(\ref{psi}), we find that the characteristic pitch-angle of plasma particles is $\psi \approx
0.4\kappa^{-0.5}$~rad. Therefore, a 'kick' that each electron gets due to the cyclotron absorption of radio waves, depends on the number of particles (determined by the Sturrock multiplication factor $\kappa$) among which the energy of these waves is distributed. For example, inside the spark-associated columns of plasma (e.g. Ruderman \& Sutherland 1975; Gil \& Sendyk 1999), where $\kappa \sim 10^{4}$, we have $\psi \sim 10^{-3}$. At the same time, in the regions of reduced number density between the spark-associated columns of plasma, where $\kappa \sim 10\div 100,$ we obtain $\psi \sim 0.1.$ For the particle to emit the synchrotron radiation the condition 
\begin{equation}
\psi \gamma_{_p}\gg 1  \label{synchro-cond}
\end{equation}
should be satisfied. Apparently, the latter condition is not met inside the
dense plasma ($\psi \gamma_{_p}\sim 1$ for $\gamma_{_p}\sim 100$), whereas in the space of reduced number density it is satisfied well. In other words, only the particles, being in the space between the spark-associated plasma columns, obtain (due to cyclotron absorption of radio waves) pitch-angles big enough to start gyrating relativistically, hence, emitting synchrotron radiation.

\section{The synchrotron model}
The total radiated power of a single electron synchrotron radiation is defined as (Ginzburg 1979; Lang 1980)
\begin{equation}
L_{0}\approx 1.6\times 10^{-15}B^{2}\gamma_{p}^{2}\sin^2\psi \ \left[ \rm{erg~s^{-1}}\right],  \label{L_0}
\end{equation}
and the critical frequency of the synchrotron radiation is 
\begin{equation}
\nu_c\approx 4.2\times 10^{6}B\gamma_{_p}^{2}\sin \psi \ \left[ \rm{Hz}\right]. \label{nu_c}
\end{equation}
The spectral density of a single electron radiation power near $\nu_c$ is 
\begin{equation}
I_{0}\left( \nu_c\right) = \frac{L_{0}}{\nu_c}\approx
3.8\times 10^{-22}B\sin \psi \ \left[ {\rm erg~s^{-1}\,Hz^{-1}}\right].  \label{I_0}
\end{equation}
Let us notice that, if a source as a whole is moving towards an observer, the power given by equation~(\ref{L_0}) should be divided by $\sin^2\psi$ (see equation (5.12) in Ginzburg (1979) and the discussion on page 75). However,
this is not the case in our current application, as the emitting volume represented in Fig.~\ref{Optics region} is quasi-stationary with respect to an observer on Earth. Indeed, the particles are permanently flowing in and out but the
volume as the whole does not approach an observer with the relativistic velocity. That is why the multiplier $1/\sin ^{2}\psi$ should be omitted in this consideration.

The average magnetic field in the region where synchrotron emission is generated (Fig.~\ref{Optics region}) can be estimated from the equation 
\begin{equation}
B=3.2\times 10^{19}\frac{\left( P\dot{P}\right) ^{1/2}}{\sin \alpha }\left( 
\frac{R_{0}}{R}\right) ^{3}=\frac{9.4\,P^{-2.5}\dot{P}_{-15}^{0.5}}
{{\mathcal R}^{3}\sin \alpha },  \label{B_Gem}
\end{equation}
as $B\approx 3.3\times 10^{3}$ G. This provides the critical frequency
$\nu_c\approx 1.4\times 10^{13}$ Hz (equation~\ref{nu_c}), which falls into infrared band (the frequency corresponding to the maximum in the single electron synchrotron spectrum yields $\nu _{m}\approx 0.29\,\nu_{c}\approx 4\times 10^{12}$ Hz). From equation~(\ref{I_0}) we find that 
$I_{0}\left( \nu_c\right) \approx 1.3\times 10^{-19}~\rm{erg~s^{-1}}~\rm{Hz^{-1}}.$ It is natural to assume that the spectral density of total power radiated by $N$ electrons is 
\begin{equation}
I\left( \nu \right) =NI_{0}\left( \nu \right) .  \label{I}
\end{equation}
Let us then estimate $N$, i.e., the total number of emitting particles. 
From Fig.~\ref{Optics region}, assuming the conservation of the particle flux along
the dipolar fields lines, we find that in the cut-cone with the height
$l=\left( {\mathcal R}_2 - {\mathcal R}_1\right) R_{_{\rm LC}}\cos\rho_1
=1.3\,R_{_{\rm LC}}\cos \rho_1\approx 1.3\times 10^{9}$ cm there are
$N\approx F_{p1}l\approx 4\times 10^{30}\kappa$ particles (where $F_{p1}$ is defined from equation~\ref{F_p}) at any given moment of time. Then, from equation~(\ref{I}), for $\kappa =10$, we obtain 
\begin{equation}
I\left( \nu_c\right) =6\times 10^{12}~\rm{erg~s^{-1}}~\rm{
Hz^{-1},}  \label{I_nuc}
\end{equation}
which is a theoretical value of spectral density near the critical infrared frequency $\nu_c$. The spectral density of power in the $B$-band can be deduced using the power-law spectrum of Geminga's optical emission (Martin et al. 1998), as
\begin{equation}
I\left( \nu_{_B}\right) =I\left ( \nu_{c}\right)\left( \frac{\nu _{_B}}{\nu_c}\right)^{-0.8} \approx 3\times 10^{11}~
\rm{erg~s^{-1}}~\rm{Hz^{-1}},  \label{I_nuB}
\end{equation}
where $\nu _{_B}=6.9\times 10^{14}$ Hz.

Let us compare the theoretical value of $I\left( \nu_{_B}\right)$ (equation~\ref{I_nuB}) with that provided by the observations. The latter can be derived from the spectral density of flux $f\left( \nu_{_B}\right) =
2\times 10^{-30}\ \rm{erg~s}^{-1}\rm{\,Hz}^{-1}$\thinspace \textrm{cm}$^{-2}$ (see Fig.~3 in Martin et al. 1998), which gives the spectral density of total power
\begin{equation}
I^{\rm{obs}}(\nu _{_B})=\varsigma \,f\left( \nu _{_B}\right) \,d^{2}\ \rm{erg~s^{-1}~Hz^{-1}} ,  \label{I_nuB_obs}
\end{equation}
where $\varsigma$ is a beaming factor and $d\approx 5\times 10^{20}$~cm. According to Fig.~\ref{Optics region}, using the features of the dipolar geometry, we find that $\phi \approx 1.5\,\rho_1\approx 0.8$ rad,
which leads to the beaming factor $\varsigma \approx \phi^{2}\approx 0.6$ sr (instead of $4\pi $ sr for the isotropic
radiation). Therefore, equation~(\ref{I_nuB_obs}) yields $I^{\rm{obs}}(\nu_{_B})\approx 3\times 10^{11}~\rm{erg~s^{-1}}~\rm{Hz^{-1}}$, which is consistent with our theoretical prediction (equation~\ref{I_nuB}). 
\begin{figure}
\epsffile[1 18 209 216]{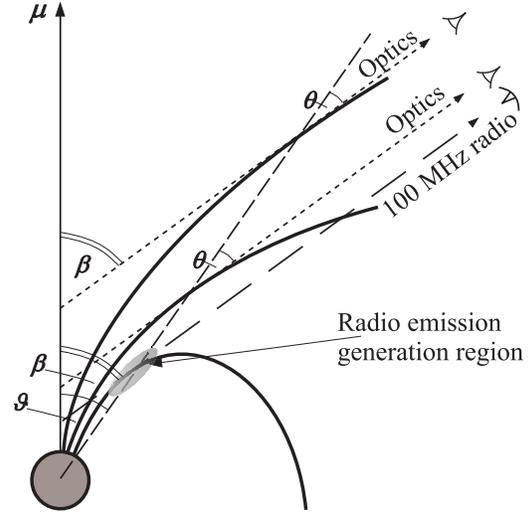}
\caption{Geometry of Geminga's optical radiation observed on Earth. $\beta$ is an impact angle of the closest approach of the observer's line of sight to the magnetic axis $\bmu$. Low-frequency radio emission which is not damped in the magnetsopheric plasma is also marked. See explanations in the text.
\label{Optics direction}}
\end{figure}

Obviously, the fraction of the synchrotron optical radiation of Geminga observed on Earth is emitted in the same direction as the low-frequency radio waves (Paper I). Using the feature of dipolar field lines that all of them intersect any given radial line with the same angle, we find a geometrical place of all the points in the magnetosphere where the tangents to the field lines point to the same direction\footnote{$\beta =20^\circ$ at the nearest approach of the line of sight to the magnetic axis $\bmu$, and $\beta+2\alpha =30^\circ$ $-$ at the furthest one.} as a wavevector ${\bmath k}\rm{_{r}}$ of the 100 MHz radio waves. As it is seen in Fig.~\ref{Optics direction}, these points are placed at the radial straight lines inclined to $\bmu$ at an angle $\vartheta \approx 13^\circ$ from the side of the nearest approach to $\bmu$ and $\vartheta\approx 20^\circ$ from the furthest one. The 'optically bright' region is restricted in the longitudinal direction to the altitudes along these lines, where the cyclotron damping of $0.1\div 1$~GHz radio waves occurs. Calculations show (Khechinashvili et al. 1999) that the longitudinal size of this region $l_{\parallel}\sim R_{_{\rm{LC}}}$. The dimension of this region in the transverse direction can be estimated as $l_{\perp}\sim R_{c}/\gamma_{p}\sim 10^{-2}R_{_{\rm{LC}}}$, where $R_{c}$ is a magnetic field line curvature at corresponding altitude. Such an estimation of $l_{\perp }$ assumes that this is a characteristic distance at which a particle moving along the curved magnetic field line keeps emitting towards the observer (i.e., an observer stays in the emission diagram of a singel particle,  within the angle range $\sim 1/\gamma_{_p}$). The corresponding characteristic time is $\tau_{\perp}\sim l_{\perp}/c\approx 4\times 10^{-4}\,\rm{s}$, during which a particle radiates $E_{0}=L_{0}\tau_{\perp }$ energy. Obviously, the latter should be much less than the kinetic energy of an electron $E_{k}=\gamma_{_p}mc^{2}$. Let us notice that in such a consideration the multiplier $1/\sin ^{2}\psi$ (see the discussion below equation~\ref{I_0}) should be taken into account in equation~(\ref{L_0}), as the source of the synchrotron radiation (a particle) is moving towards the observer in this case. Using equation~(\ref{L_0}) with the latter correction we find that $\chi \equiv E_{0}/E_{k}\sim 10^{-3}\ll 1,$ so that only a tiny fraction of the particle's kinetic energy is radiated away as a synchrotron emission. Assuming that this is valid for all the radiating particles, we
can estimate the total radiated power of the synchrotron radiation as $L=\chi L_{p1}\approx 8\times 10^{25}~{\rm erg~s^{-1}}$ (where $L_{p1}$ is found from equation~\ref{L_p}). Then, using equation~(\ref{I_0}),
we obtain $I\left( \nu_c\right)\approx 6\times 10^{12}~\rm{erg~s^{-1}}~\rm{Hz^{-1}}$. We see that this value is consistent with our previous theoretical estimation (equation~\ref{I_nuc}), and hence with the observationally derived value (equation~\ref{I_nuB_obs}). The consideration based on the energy conservation provides an independent test of our model.

In the calculations above we assume that both pitch-angles $\psi$ and Lorentz factors $\gamma_{_p}$ of emitting particles remain constant while the particles cross the 'optically bright' region. Obviously, this assumption is valid for Lorentz factors because $\chi \ll 1$. As for pitch-angles, it requires that a single particle goes through at least few acts of radio-absorption during the time $\tau_{\rm m}=l_{\parallel}/c\approx 0.04$~s. This is equivalent to fulfillment of the following condition
\begin{equation}
\tau_{d}/\tau_{\rm m}\ll 1.  \label{tau_d/tau_m}
\end{equation}
Here $\tau_{d}= 1/\Gamma$ is a characteristic time-scale of the cyclotron damping, and $\Gamma$ is the damping decrement. We found (Paper~I; Khechinashvili et al. 1999) that $\Gamma R/c\sim 100$ at the distances where the bulk of radio band ($0.1\div 1$ GHz) is typically damped. Therefore, $\Gamma \sim 2\times 10^{3}\,{\rm s^{-1}}$ and $\tau_{d}/\tau _{\rm m}\sim 10^{-2}\ll 1$. Thus, the parameters $\gamma_{_p}$ and $\psi$ used in our model can be considered quasi-stationary.

We believe that the apparent optical spectral index $\alpha\approx 0.8$ (Martin et al. 1998) indicates a power-law distribution of emitting particles over energy $n\left( \gamma \right) \sim \gamma^{-q}$, where $q=1+2\alpha \approx 2.6$ (Ginzburg 1979; Lang 1980). Such a distribution function can be expected in the region where the synchrotron optical radiation of the Geminga pulsar is believed to originate.

\section{Conclusions}
The model for the optical radiation of the Geminga pulsar, presented in this paper, is a direct consequence of our model for its erratic radio emission (Paper I). Let us summarise our main results:  (i) Relativistic charged particles of the pulsar magnetosphere obtain non-zero pitch-angles due to cyclotron damping of radio waves; (ii) The particles re-emit the absorbed radio energy in the form of synchrotron radiation, with a maximum power in the infrared band; (iii) The observed optical emission of the Geminga pulsar is just a short-wavelength continuation of its power-law synchrotron spectrum; (iv) The optical power estimated from our model, using two different approaches, agrees with the observed value; (v) We predict a significant pulsed infrared emission which should be observed from the Geminga pulsar; (vi) As a by-result, we support a model of the non-uniform plasma outflow in pulsar magnetospheres, originating in the form of spark discharges above the polar cap.
\section{Acknowledgments}
The work is supported in parts by the KBN Grants 2 P03D 015 12 and 2 P03D 003 15 of the Polish State Committee for Scientific Research. G. M. and D. K. acknowledge also support from the INTAS Grant 96-0154.

\label{lastpage}

\end{document}